\documentclass{elsart5p}

\usepackage{natbib}
\usepackage{graphicx}
\usepackage{amssymb}

\begin{document}
\begin{frontmatter}

\title{Tailor-made Quantum Well-in-a-Well Systems: \\Their
Bound States and Scattering Properties}

\author{N.J. Jacosalem} and
\author{C. Villagonzalo}
\ead{cvillagonzalo@nip.upd.edu.ph}

\address{Structure and Dynamics Group, National Institute of Physics,\\
University of the Philippines, Diliman, Quezon City 1101, Philippines}

\begin{abstract}
The resulting stationary states and scattering properties of an effective potential
brought about by embedding a quantum well in another well are  investigated in this work. 
The composite well system is constructed via a superposition of modified
P\"{o}schl - Teller potential wells.  The energy spectrum in each composite well  
is obtained using the shooting method and the transport of a particle above this system
is analyzed using the transfer matrix method.
It is shown that decreasing the size of the embedded middle well lowers
the ground state energy of the well-in-a-well system. Moreover, the bound states increase 
in number and become more evenly spaced. In addition, the transmission probability 
of a free particle incident above a composite well is lowest for 
the system with a large embedded well as compared to well-in-a-well systems 
of the same depth. Small variations in designed potential wells
yield different quantum mechanical features.
\end{abstract}


\end{frontmatter}
\section{Introduction}
\label{sec1}

The search for increased functionality in semiconductor-based devices
is made possible with the advancement of microfabrication and 
epitaxial growing techniques. A quantum well, built from two wider-bandgap 
semiconductors separated by a thin layer of narrower-bandgap semiconductor,
can now be designed to deviate from the conventional rectangular- or 
square-well profiles in order to obtain more efficient properties. 
A study has shown, for example, that $n$-doped parabolic quantum wells absorb 
far-infrared radiation at the bare-harmonic-oscillator frequency independent 
of electron-electron interactions and the number of electrons in the 
well \cite{BreyJH89}.
Another device is that of a heterostructure made from a high bandgap 
"spike" placed in  the middle of a rectangular quantum well which showed a reduced 
material gain leading to an increased threshold current \cite{LaaksoDTT07}.
Furthermore, simulations on a diode laser based on strained non-square shaped 
quantum well yield enhanced radiative current performance as compared to
a device based on an optimal square well of the same width and emission length
\cite{KadukiGBA03}. The authors  of Ref.~\cite{KadukiGBA03} concluded that their 
embedded quantum well design may be suitable for optical confinement and carrier 
capture.

With the advances in band-gap engineering, it is suitable to have
an easily manipulated quantum well model that yields the optimized properties 
prior to fabrication. 
A technique was developed using supersymmetric quantum mechanics to optimize the 
quantum well structure in respect to maximizing the gain in optically pumped 
intersubband lasers \cite{TomicMZ00,RadovanovicMII99}. 
This method adds a bound state lower than the existing ground state energy of
a potential well, thereby, varying the well's initial shape in the process. 
The resulting quantum well may not have the symmetric structure of the initial well used.
In contrast, this work will show that one obtains a lower ground state by
an appropriate embedding of a quantum well in another quantum well while maintaining
the symmetry of the initial composite potential. 

Here a composite quantum well is constructed 
through the use of modified P\"{o}schl-Teller (MPT) potentials \cite{Flugge74} similar 
to that used in Ref.~\cite{TomicMZ00}. These potentials are
related in form to Rosen-Morse potentials \cite{KleinertM92,RosenM32}
and have been used successfully to model disordered quantum wires \cite{Rodriguez06}.
The MPT-type of potentials are chosen since they offer a high degree of control 
and flexibility.
Moreover, different single potential  wells can be joined smoothly at the edges forming 
one continuous potential. Therefore, the systematic numerical procedure established
in solving the eigenvalue equation for one composite quantum well can readily be used
even when the parameters of the constituent single wells are varied.

The effective changes in the energy spectrum and scattering
properties that occur when a quantum well is embedded in another well
will be investigated in this paper. This will serve as aid to experiments in that
constructed composite quantum wells of the same type and symmetry
with different embedded well sizes have fundametally different features.

\begin{figure}[t]
\centering
\includegraphics[width=8.2cm]{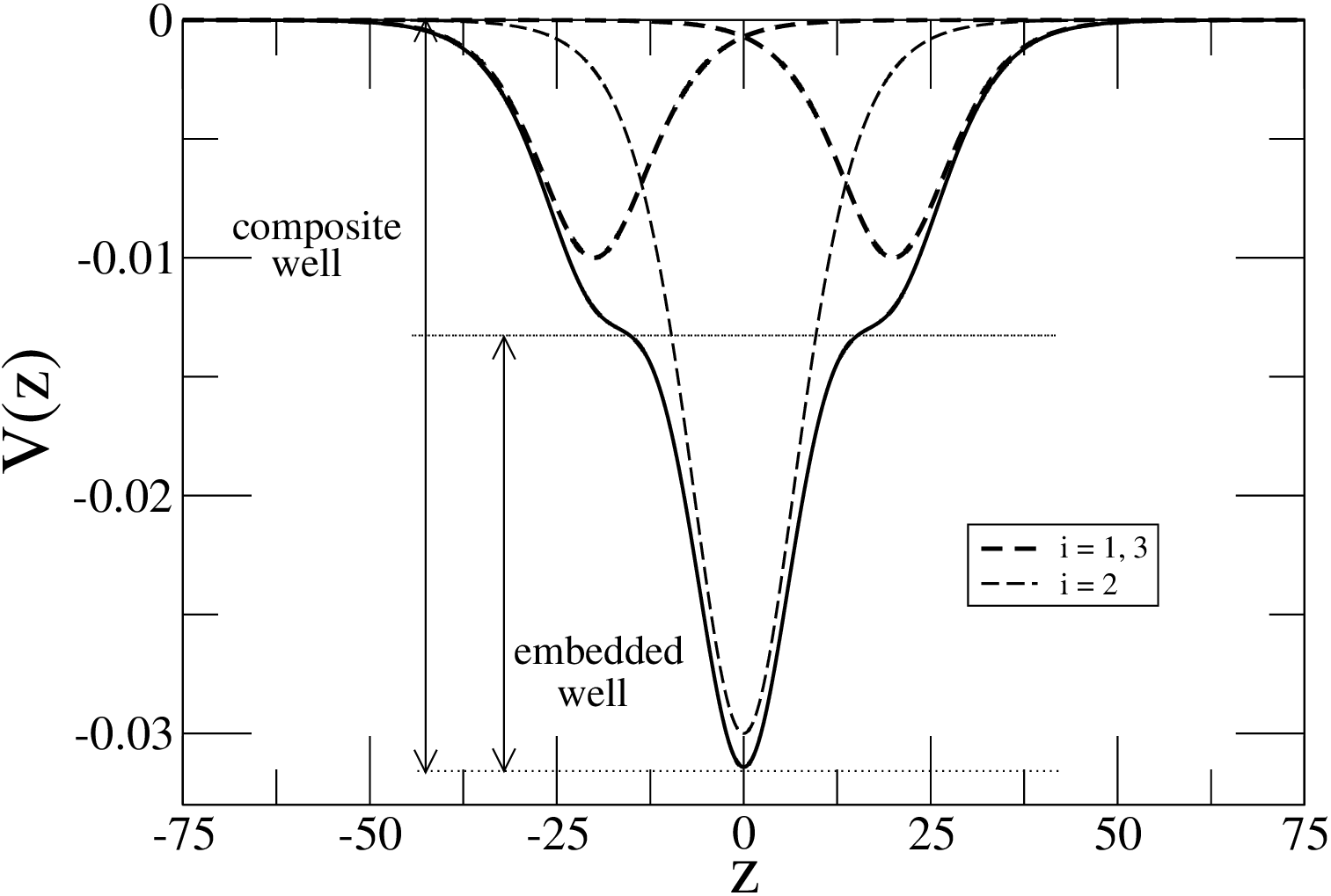}
\vspace{0.2cm}
\caption{A sum of 3 MPT potentials yields an embedded quantum well
in another well. Here $V(z)$ is in units of $\hbar^2/m^{*}$ and $z$ is in
angstrom.}
\label{fig0}
\end{figure}

\section{The Quantum Well-in-a-Well Model}
\label{sec2}

A quantum well-in-a-well can be constructed from a sum of three MPT potentials, that is,
\begin{equation}
V(z) = -\left(\frac{\hbar^2\alpha^2}{2m^*}\right) \sum_{i=1}^{3}  
\frac{\lambda_i(\lambda_i - 1)}{\cosh^2 [\alpha (z+d_i)]}\;,
\label{potential}
\end{equation}
where $m^*$ is the effective mass and $\hbar$ is Dirac's constant.
The well-in-a-well system consists of two left and right wells, labelled with
indices $i=1$ and $i=3$, respectively, and a center well denoted by index $i=2$. 
Their location relative to the origin are  $d_1= -d$, $d_2 = 0$, and $d_3=+d$.
In this work, the width parameter $\alpha$ is the same for all wells studied
and it is kept to a constant value of $0.1\;\AA^{-1}$. Only the value of the depth
parameter $\lambda$ is varied. Here the depth parameters of the side wells, 
$i_1\;\;\&\;\;i_3$,  are set equal, that is, $\lambda_1=\lambda_3$. This is done 
to retain the symmetric shape of the well about the origin. The middle well has a 
depth parameter $\lambda_2$ and its value can be different from $\lambda_1$.
By varying the values of $\lambda_1$, $\lambda_2$ and the shift parameter $d$, 
the size of the effective embedded potential well relative to the resulting main quantum 
well can be controlled. Figure 1 illustrates a composite well as obtained from its constituent 
potential wells. Note that the model represents the conduction band of a quantum well system.

\section{The Numerical Method}

The bound states of a single electron in the composite well system described above 
can be obtained by using Eq.~(\ref{potential}) in the Schr\"{o}dinger equation
in one dimension
\begin{equation}
\left[-\frac{\hbar^2}{2m^*}\frac{\partial^2}{\partial z^2}+V(z)\right]\psi(z)=E\psi(z)\;.
\label{SE}
\end{equation}
It follows that the wavefunction can be determined through an iterative 
procedure from Eq.~(\ref{SE}) in the central difference form \cite{Harrison00}, that is,
\begin{eqnarray}
\psi(z+\delta z) = \frac{2m^*}{\hbar^2}(\delta z)^2[V(z) - E]\psi(z) 
\nonumber \\
+2\psi(z) 
- \psi(z-\delta z)\;.
\label{wavefunction}
\end{eqnarray}
Here $\delta z$ is an arbitrary infinitesimal step size and the initial values of
$\psi(z)$ and $\psi(z-\delta z)$ are obtained via simple symmetry arguments. 
The advantage of using the MPT potentials is that the eigenenergies of a single MPT potential
is known analytically \cite{Flugge74}. Hence, the difference in energy states of the constructed composite well as compared to the single MPT well can be related to the potential parameters.

The shooting method \cite{Harrison00} is implemented in this work to obtain each energy 
eigenvalue in a given energy range. Each solution must satisfy the boundary conditions that
 $\psi(z)$ and its derivative vanish at infinity. In addition to the boundary conditions, the minimum tolerance set for numerical convergence of $\psi$ and its derivative is $10^{-10}$.
To accurately obtain the bound states, the whole potential depth is scanned to check for
energy intervals wherein the first derivative 
of the wavefunction changes sign at infinity. 
This signals that within this energy range a bound state can be found. 
The bisection method is then
applied to this particular interval to search for the bound state with a convergence 
limit of  $\Delta E = 10^{-10} \hbar^2/m^*$.

Another property of this well-in-a-well system that will be studied here is the
scattering of a free particle from this potential landscape via the transfer matrix approach.
The transmission probability is obtained from the ratio between 
the amplitude of the transmitted wave ($A_T$) and that of the incident wave ($A_I$)
\begin{equation}
|T|^2 = \left(\frac{|A_T|}{|A_I|}\right)^2 = \frac{1}{|M_{11}|^2}\;.
\end{equation}
The transfer matrix technique yields \cite{Rodriguez06,Singh97}
\begin{equation}
M_{11}=\frac{1}{2}\left[1+\frac{\kappa(z+\delta z)}{\kappa(z)} \right] e^{i[\kappa(z+\delta z)
-\kappa(z)]z}\;,
\end{equation}
where 
$\kappa(z) = \left(2m^*[E-V(z)]\right)^{1/2}/\hbar$. Here $E$ is the incident
particle's kinetic energy. It is further assumed here that the effective 
mass does not vary in space. Length measurements are given in angstroms and 
energy measurements are in units of ${\hbar^2}/{m^*}$.

\section{Results and Discussion}
\label{sec3}

\begin{figure}[t]
\centering
\includegraphics[width=8.2cm]{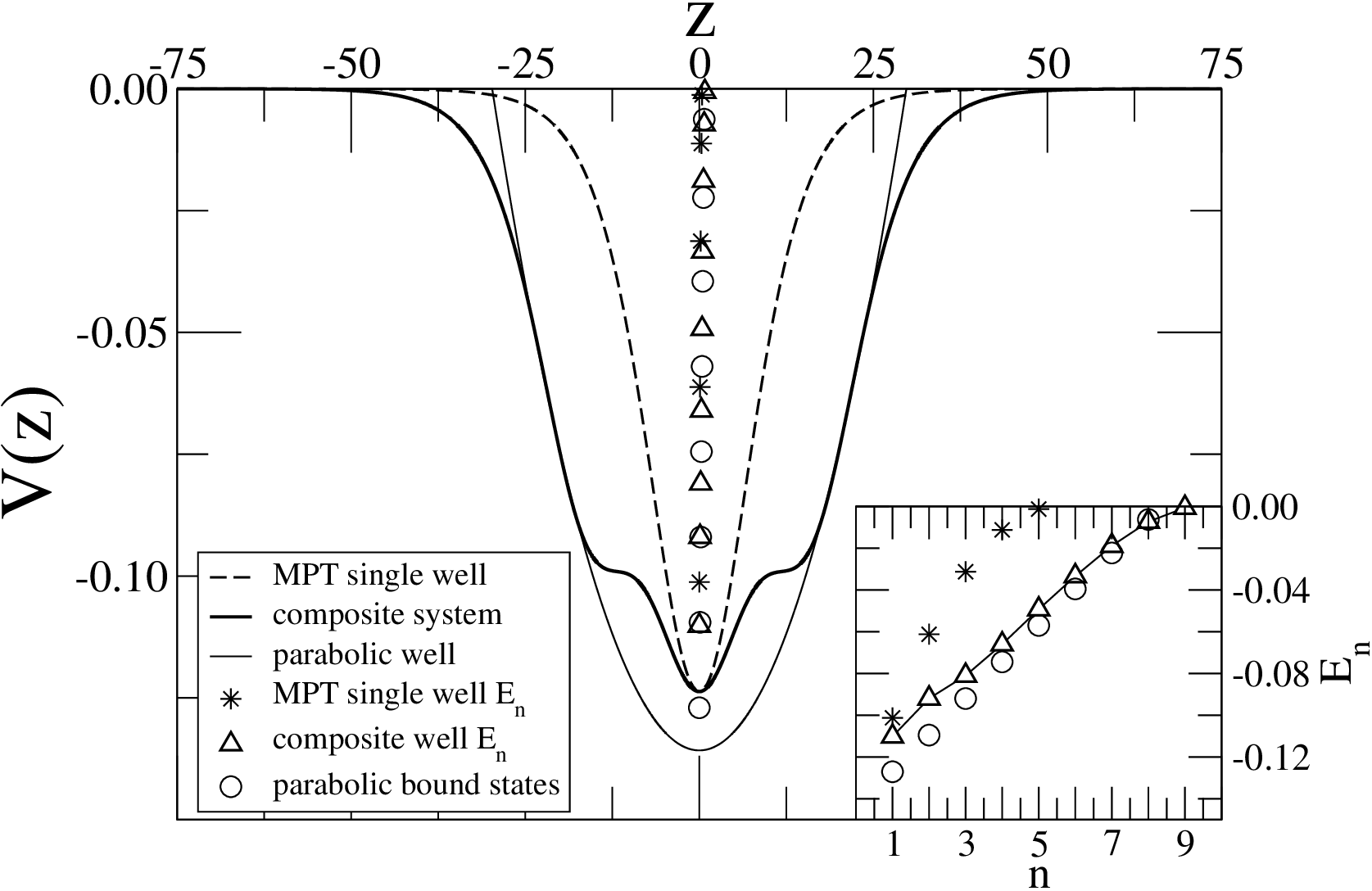}
\vspace{0.2cm}
\caption{Bound states of a composite quantum well-in-a-well system as compared 
to a single MPT well of the same depth and a parabolic well of the same width. 
For the composite system, the constituent well parameters are 
$\lambda_1=\lambda_3=4.5$, $\lambda_2=5.0$ 
and $d=16\;\AA$, while for the single MPT well $\lambda = 5.4992$. The inset shows the 
eigenenergies $E_n$ where $n$ is the bound state index.}
\label{fig1}
\end{figure}

Figure \ref{fig1} shows a quantum well-in-a-well system, whose 
subwell size is 24$\%$ of the full well depth.
The properties of the composite well is studied using a single MPT well of the same depth
as a benchmark. In addition, a parabolic well having the same width is fitted to the composite well for comparison. The same approach is used to obtain the bound states
for the single MPT well and the parabolic well. 

One finds that the composite well's ground state is lower by
$\approx 9\%$ relative to the ground state of the single MPT well of the same depth.
In this example, the ground state of the parabolic well is lowest as expected due to 
its deeper potential depth. 
Note that the composite well has more bound states than that of single
MPT well of the same depth or the parabolic well of the same width. 
Further, the interval between bound states are more evenly distributed in the composite
system as compared to the single MPT well. The distribution of states for 
the composite well is more similar to the even distribution of energy states of a 
parabolic well of the same width rather than that of the single MPT potential. 

\begin{figure}[t]
    \centering 
    \includegraphics[width=7.8cm,height=!]{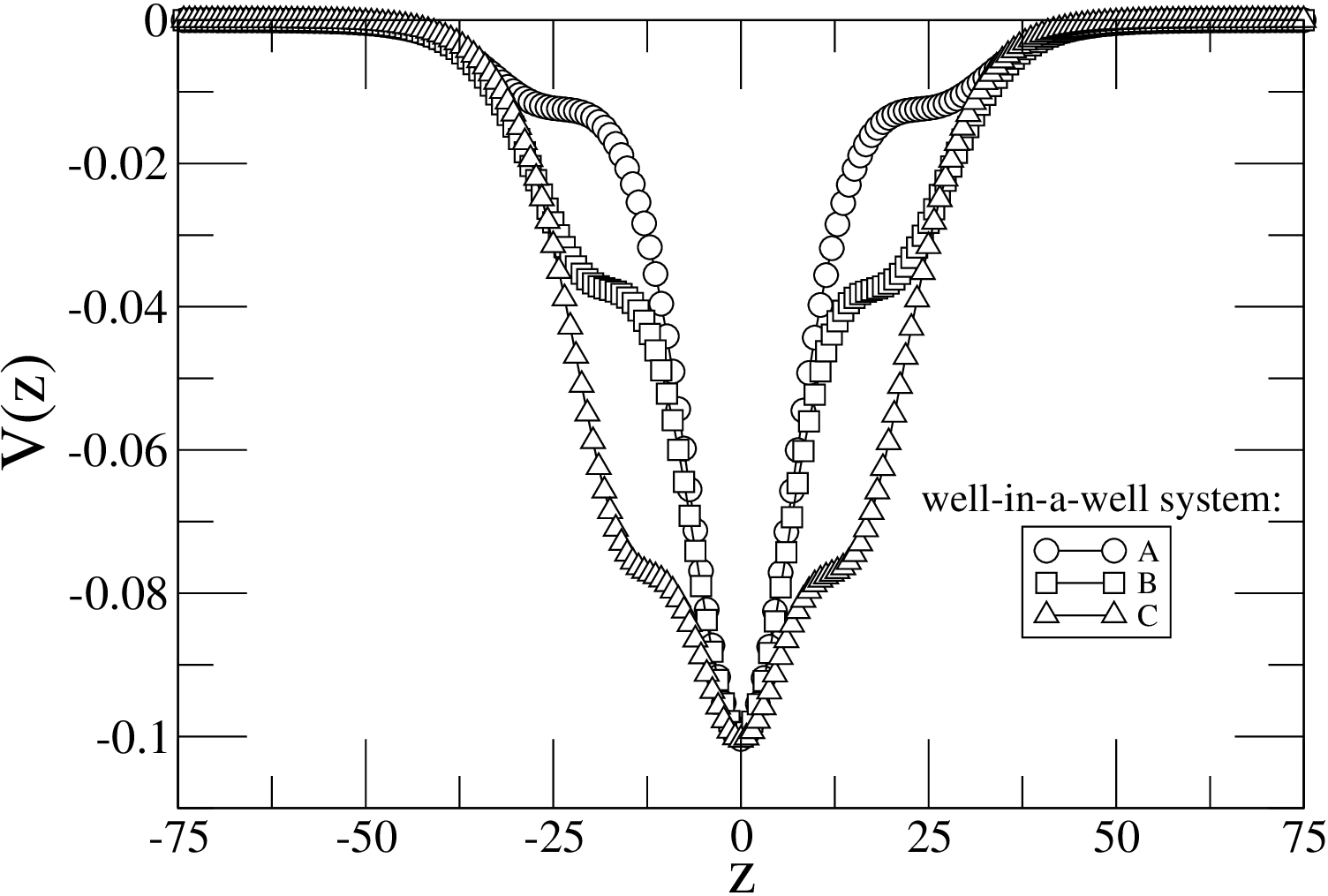}
\vspace{0.2cm}
    \caption{Three quantum well-in-a-well systems having equal depth and 
base width. 
The other potential parameters are (A.) $d=27\;\AA$, $\lambda_1 = \lambda_3 =2$ and 
$\lambda_2 = 5$, (B.) $d=21\;\AA$, $\lambda_1 =\lambda_3 = 3.00$ and $\lambda_2 = 4.92$, 
and (C.) $d=16\;\AA$, $\lambda_1  =\lambda_3 = 4.00$ and $\lambda_2 = 4.59$.}
    \label{fig3}
 \end{figure}

Next, Fig.~\ref{fig3} illustrates three well-in-a-well systems 
of the same depth, symmetry and base width.
The latter is  the potential width at $V(z)=0$. What varies in this plot 
is the depth of the component side wells yielding systems having different 
depths of embedded wells.
Figure \ref{fig3} shows systems with  embedded wells of depths
(A) 88$\%$, (B) 64$\%$ and (C) 28$\%$ relative to the composite well's full depth.

\begin{figure}[b]
\centering
    \includegraphics[width=7.8cm,height=!]{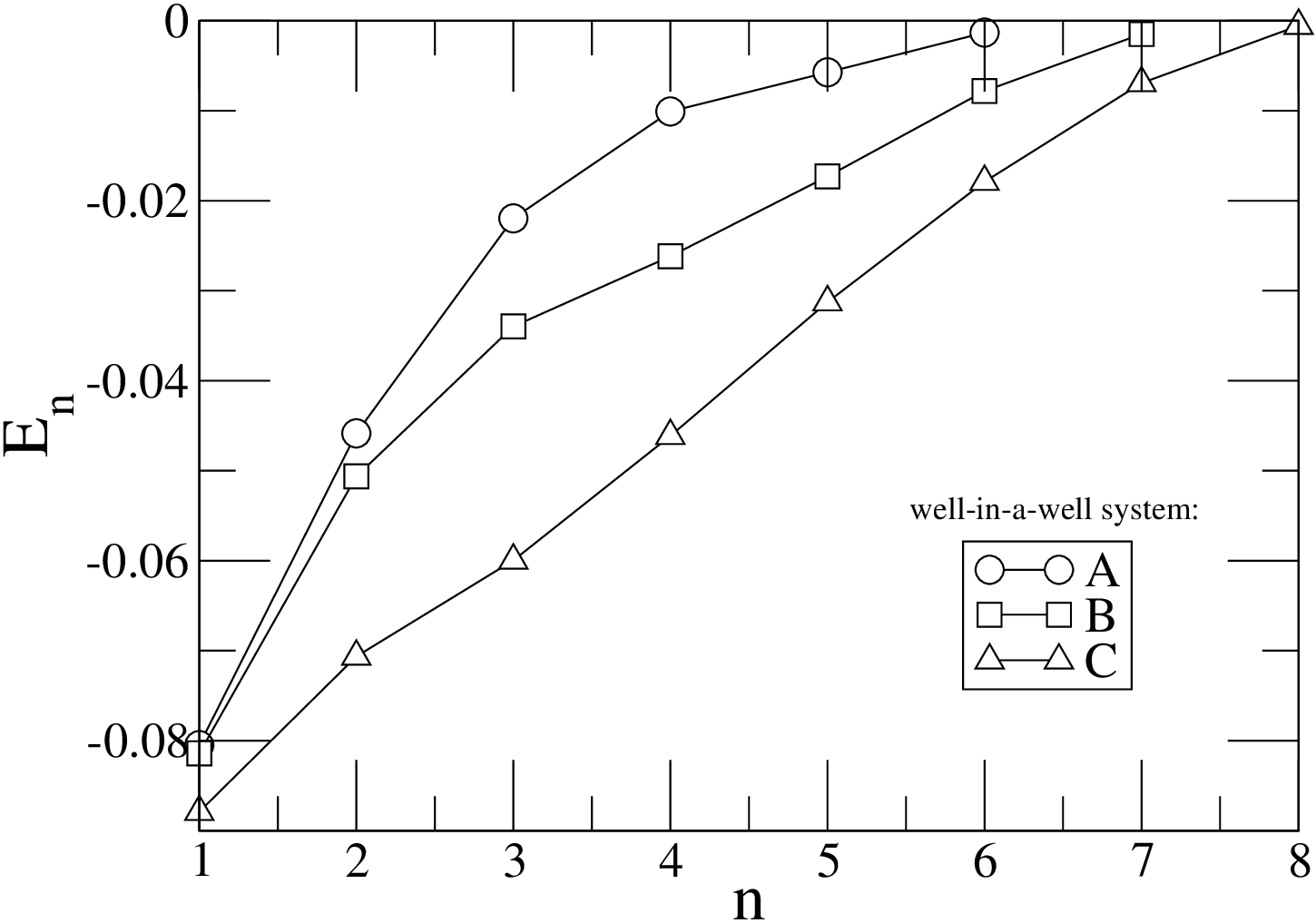}
\vspace{0.05cm}
    \caption{The bound states of the three well systems
shown in Fig.~\ref{fig3}.}
    \label{fig4} 
 \end{figure}

The corresponding bound states of the three composite wells are shown in
Fig.~\ref{fig4}. As the depth of the component side wells increases the 
number of bound states also increases. Thereby, well (C) has the most number 
of bound states.
Furthermore, the change in slope of the plot of eigenstate $E_n$ relative to the 
bound state index $n$, towards a constant value indicates the shift towards 
an even distribution of the energy states. 
This occurs for the case of well (C) since the shape of the edge of this well
approaches that of a parabolic potential. Recall that for the case of the simple harmonic
potential well, the energy levels are evenly distributed. So as the bottom of the 
component side wells approaches the depth of the middle well, we expect a straight line.
However, near the base of the potential the shape retains the tail of a modified
P\"{o}schl-Teller potential well, hence the last two bound states are nearer 
to each other as compared to those adjacent states in the middle and edge of the well. 

Another phenomenon which is affected by the potential is
the transport of a particle above it. Recall that
in the case of a finite rectangular well of width $L$, or similarly, a potential barrier
of the same width, transmission resonance is observed
when the wave number takes on integral multiples of $\pi/L$ \cite{Harrison00}.
It has also been demonstrated that a low energy incident electron above a 
rectangular well may be captured into a bound state due to dissipation \cite{CaiHZY90}.
Transmission resonance is only restored for particle kinetic energies beyond the "captive" 
energy region \cite{CaiHZY90}. 

\newpage

\begin{figure}[t]
\centering
   \includegraphics[width=8.2cm,height=!]{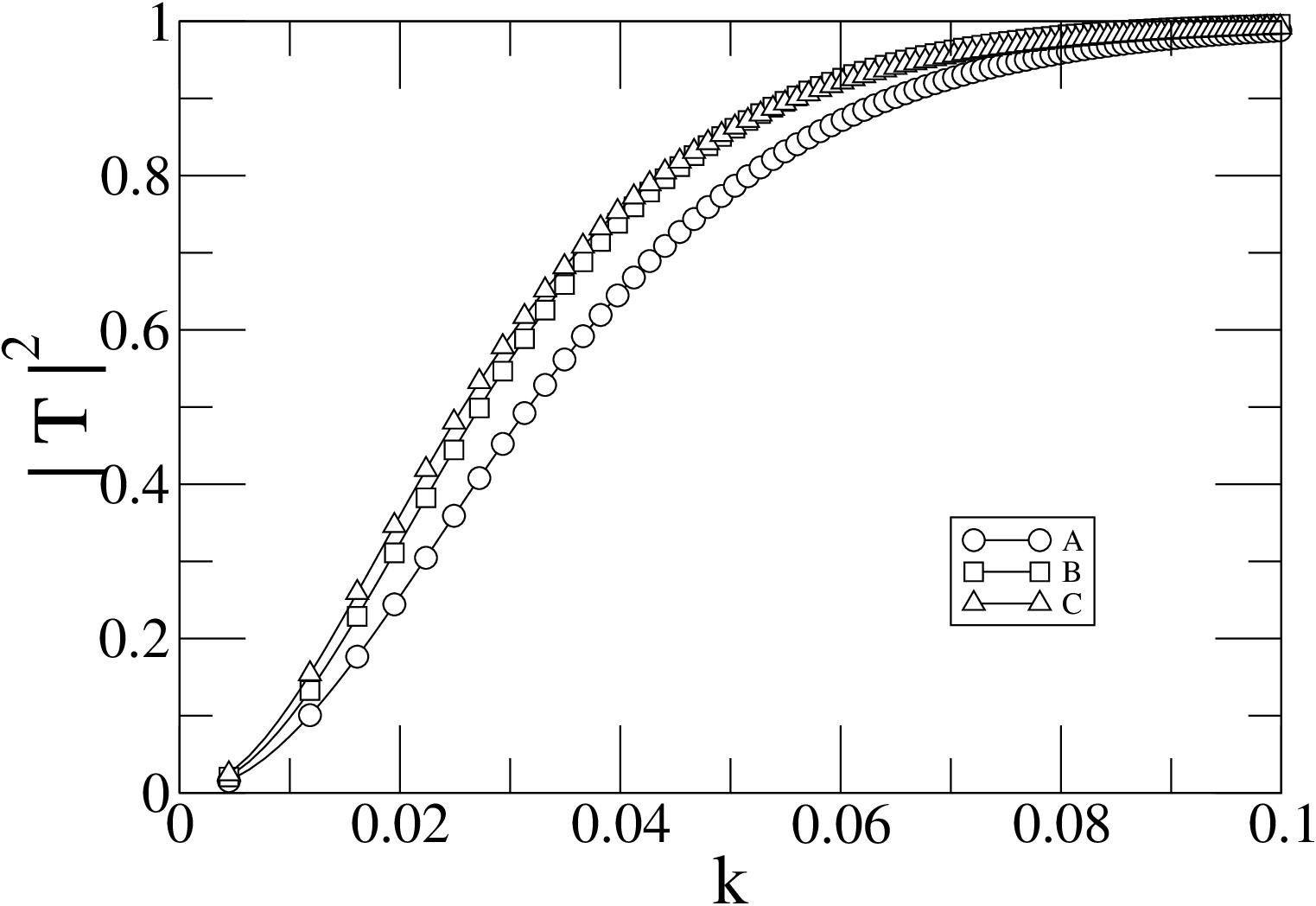}
    \caption{The corresponding transmission probability 
of an incident free particle, with energy $E>0$, scattered on each 
well as shown in Fig.~\ref{fig3}.
 Here $k$ is the magnitude of the free particle's wave vector
and is given in units of $(m^{*})^{1/2}/\hbar$.}
    \label{fig5} 
 \end{figure}

In the case of scattering above a single MPT potential, 
transmission resonance is observed when $\lambda$ is an integer 
regardless of the magnitude of the kinetic energy of the incident
particle \cite{Flugge74}. Unlike in the well-in-a-well systems
studied here, there are no resonant wells. This is true even 
for well (A) in which its constituent wells, by themselves,
are absolute transparent potentials. 
Figure \ref{fig5} shows the transmission probability of a free particle
with effective mass $m^*$  above each composite well in Fig.~\ref{fig3}. 
A particle has a lower probability
of transmission  if the kinetic energy of the particle approaches zero. 
As expected, the larger the kinetic energy of the incident particle
the more likely it is to be transmitted.
The monotonic increasing trend of the transmission probability 
for the composite wells remains valid even when the well width is increased.
This is in contrast to the appearance of an oscillatory nature of
the transmission coefficient when the width of a finite rectangular
well is widened.

The behavior of interest is that well (A) has the lowest probability
of transmission relative to systems (B) and (C). 
Note that for a particle scattered in a finite rectangular well 
in the non-resonance regime, the 
transmission probability increases with decreasing depth and width.
The opposite behavior is, thus, observed here, wherein the most
shallow side wells and the most narrow middle well yield a composite
system that creates more disturbance to particle transmission.
In the perspective of an incident particle with energies
corresponding to $0.2 (m^{*})^{1/2}/\hbar <k<0.6 (m^{*})^{1/2}/\hbar$,
wells (B) and (C) are more slowly varying potential functions
relative to well (A). The abruptness of the change in the potential in 
(A) provides a stronger force in reducing the probability of transmission
in this $k$ regime.

\section{Summary}

This paper presented a simulation model for a composite quantum well-in-a-well
system and investigated the quantum mechanical properties arising from such 
construction. A superposition of modified P\"{o}schl-Teller potentials is chosen 
for the model because the constituent wells can easily be varied
without increasing the numerical complexity in
solving the Schr\"{o}dinger equation.
This work showed that deviations in quantum well structures of the same
functional form and depth yield  entirely
different properties as shown in the differences in their ground state energies, 
the number and distribution of bound states and the transmission probabilities 
of an incident free particle above the composite wells.
Tailor-made quantum well systems as presented 
here offer ease and flexibility that can be suited to desired features
for application purposes.

\section*{Acknowledgment}
C. Villagonzalo is grateful for the support provided by the
Office of the Vice President for Academic Affairs  through the
University of the Philippines System Grant. 

\bibliographystyle{plain}

\end{document}